\documentclass[a4paper]{article}
\usepackage{amsmath, epsfig}
\newcommand{\pdif}[2]{\ensuremath{\frac{\partial #1}{\partial #2}}}
\newcommand{\pdiftwo}[2]{\ensuremath{\frac{\partial^2 #1}{\partial
#2^2}}} 

\title{Comparison of Field Theory Models of Interest Rates with Market Data}
\author{Belal E. Baaquie and Marakani Srikant\\Department of Physics, National University of Singapore\\phybeb@nus.edu.sg, srikant@srikant.org}
\begin{document}
\maketitle
\begin{abstract}
  We calibrate and test various variants of field theory models of the
  interest rate with data from eurodollars futures. Models based on
  psychological factors are seen to provide the best fit to the
  market. We make a model independent determination of the volatility
  function of the forward rates from market data.
\end{abstract}

\section{Introduction}
In this paper, we compare field theory models of interest rate models
with market data, and propose certain modified models inspired from
theoretical considerations and observed facts about the interest
rates. The theoretical framework for all these models is Baaquie's
formulation \cite{Baaquie}, \cite{Baaquiesv} of forward rates as a two
dimensional quantum field theory. The Baaquie model is a
generalization of the Heath-Jarrow-Morton (HJM) model; the key feature
of the field theory model is that the forward rates $f(t,x)$ are
imperfectly correlated in the maturity direction $x>t$, and which is
specified by a rigidity parameter $\mu$. The models we study are the
following: (a) forward rates with constant rigidity \cite{Baaquie},
(b) forward rates with the variation of the spot rate constrained by a
new parameter \cite{Baaquienotes}, and two new models proposed in this
paper, namely (c) forward rates with maturity dependent rigidity
$\mu(x-t)$, and lastly (d) forward rates with non-trivial dependence
on maturity specified by an aribitrary function $z=z(x-t)$.

We first briefly review Baaquie's
field theory model and review the market data used in this study. We
then test the field theory model, introduce two variants and test them
as well. We find that the correlation structure can be explained by a
relatively straightforward two parameter model which also has a useful
theoretical interpretation. 

\section{The HJM model}
\subsection{Definition of the model}
In the HJM- model the forward rates are given by 
\begin{equation}
  \label{eq:hjmdef}
  f(t, x) = f(t_0, x) + \int_{t_0}^t dt' \alpha(t', x) + \sum_{i=1}^K \int_{t_0}^t
  dt' \sigma_i(t',x) dW_i(t')
\end{equation}
where $W_i$ are independent Wiener processes. We can also write this
as 
\begin{equation}
  \label{eq:hjmdefint}
  \pdif{f(t,x)}{t} = \alpha(t,x) + \sum_{i=1}^K \sigma_i(t,x) \eta_i(t)
\end{equation}
where $\eta_i$ represent independent white noises. The action
functional, is 
\begin{equation}
  \label{eq:hjmaction}
  S[W] = -\frac{1}{2}  \sum_{i=1}^K \int dt \eta_i^2(t)
\end{equation}
We can use this action to calculate the generating functional which is
\begin{eqnarray}
\label{eq:z}
Z[j,t_1,t_2]&=&\int {\cal D}W e^{\sum_{i=1}^K\int_{t_1}^{t_2}dt
    j_i(t)W_i(t)}e^{S_0[W,t_1,t_2]}\nonumber  \\
    &=&e^{\frac{1}{2}\sum_{i=1}^K \int_{t_1}^{t_2}dt j_i^2(t)}
\end{eqnarray}

\section{Field theory model with constant rigidity}
We now review Baaquie's field theory model presented in \cite{Baaquie}
with constant rigidity. Baaquie proposed that the forward rates being
driven by white noise processes in (\ref{eq:hjmdefint}) be replaced by
considering the forward rates itself to be a quantum field. To
simplify notation, we write the evolution equation in terms of the
velocity quantum field $A(t, x)$, and which yields
\begin{equation}
  \label{eq:baaquiedef}
  f(t, x) = f(t_0, x) + \int_{t_0}^t dt' \alpha(t', x) + \sum_{i=1}^K \int_{t_0}^t
  dt' \sigma_i(t',x) A_i(t',x)
\end{equation}
or 
\begin{equation}
  \label{eq:baaquiedefdif}
  \pdif{f(t,x)}{t} = \alpha (t,x) + \sum_{i=1}^K \sigma_i(t,x) A_i(t,x)
\end{equation}
The main extension to HJM is that $A$ depends on $x$ as well as $t$
unlike $W$ which only depends on $t$. 

While we can put in many fields $A_i$, we will see that the extra
generality brought into the process due to the extra argument $x$ will
make one field sufficient. Hence, in future, we will drop the
subscript for $A$. 

Baaquie further proposed that the field $A$ has the free (Gaussian) free  field action functional 
\begin{equation}
  \label{eq:adef}
  S = -\frac{1}{2} \int_{t_0}^{\infty} dt \int_t^{t+T_{FR}} dx \left(A^2 +
    \frac{1}{\mu^2} \left(\pdif{A}{x}\right)^2\right)
\end{equation}
with Neumann boundary conditions imposed at $x=t$ and
$x=t+T_{FR}$. This makes the action equivalent (after an integration
by parts where the surface term vanishes) to 
\begin{equation}
  \label{eq:adef1}
  S = -\frac{1}{2} \int_{t_0}^\infty dt \int_t^{t+T_{FR}} dx A(t, x)\left(1
    - \frac{1}{\mu^2}\pdiftwo{}{x}\right)A(t, x) 
\end{equation}
This action has the partition function 
\begin{equation}
  \label{eq:partition}
  Z[j] = \exp\left(\int_0^{t_1} dt \int_t^{t+T_{FR}} dx dx' j(t,x)
    D(x-t,x'-t) j(t,x')\right)
\end{equation}
with 
\begin{equation}
  \label{eq:D}
  \begin{split}
    D(\theta,\theta';T_{FR})&=\mu\frac{\cosh\mu\left(T_{FR}-|\theta-\theta'|\right) + \cosh\mu\left(T_{FR}-(\theta+\theta')\right)} {2\sinh\mu T_{FR}}\\   
    &= D(\theta',\theta;T_{FR})\quad : \quad
    \mathrm{Symmetric~Function~of}~\theta,\theta'
\end{split}
\end{equation}
where $\theta = x-t$ and $\theta' = x'-t$. We can calculate
expectations and correlations using this partition function.  Note
that due to the boundary conditions imposed, the inverse of the
differential operator $D$ actually depends only the difference $x-t$.
The above action represents a Gaussian random field with covariance
structure $D$. In \cite{Baaquie}, a different form was found as the
boundary conditions used were Dirichlet with the endpoints integrated
over. This boundary condition is in fact equivalent to the Neumann
condition which leads to the much simpler propagator above. In the
limit $T_{FR} \rightarrow \infty$ which we will usually take, the
propagator takes the simple form $\mu e^{-\mu \theta_>} \cosh \mu
\theta_<$ where $\theta_>$ and $\theta_<$ stand for $\max(\theta,
\theta')$ and $\min(\theta, \theta')$ respectively.

When $\mu \rightarrow 0$, this model should go over to the HJM model.
This is indeed seen to be the case as it is seen that $\lim_{\mu
  \rightarrow 0} D(\theta, \theta'; T_{FR}) = \frac{1}{T_{FR}}$. The
extra factor of $T_{FR}$ is irrelevant as it is due to the freedom we
have in scaling $\sigma$ and $D$. The $\sigma$ we use for the
different models are only comparable after $D$ is
normalized\footnote{This freedom exists since we can always make the
  transformation $\sigma(\theta) \sim \eta(\theta) \sigma(\theta)$ and
  $D(\theta, \theta') \sim D(\theta, \theta')/(\eta(\theta)
  \eta(\theta'))$ without affecting any result}. On normalization, the
propagator for both the HJM model and field theory model in the limit
$\mu \rightarrow 0$ is one showing that the two models are equivalent
in this limit.

The basic model with constant rigidity can be generalized in many
different ways. The generalization to positive valued forward rates,
and to models with stochastic volatility are studied in
\cite{Baaquiesv}. In this paper we generalize the free field model to
more complex dependence of $\mu$ and $f(t,x)$ on the maturity
direction $\theta$.

\section{The Market Data used for the Study}
We used the Eurodollar futures data for the following study. A
Eurodollars futures contract is represents a deposit of US\$1,000,000
for three months at some time in the future. Currently, futures
contracts for deposits upto ten years into the future are actively
traded. Significant historical data for contracts on deposits upto
seven years into the future are available. If one makes the reasonable
approximation that $f(t, \theta)$ is linear for $\theta$ between
contract times, one can use this data as a direct measure of the
forward rates. 

Further, the straightforward simplification that the
Eurodollar futures prices directly reflect the forward rate was done,
an assumption previously used in the literature \cite{data1}. This is
a reasonable assumption as the forward rates are small enough that the
difference between the logarithmic measure of the forward rate used in
theory and the arithmetic rates used in the market are insignificant.
We also attempted to analyse Treasury bond tick data from the GovPx
database but we found it impossible to obtain forward rates accurate
enough for our purposes. The main reason for this is that while we
were able to obtain reasonably accurate yields for a few maturities,
the differentiation required to get the forward rates from the yields
introduced too many inaccuracies. This is somewhat unfortunate since
Treasury bonds represent risk free instruments while a small credit
risk exists for Eurodollar deposits.

For the following analysis, we used the closing prices for the
Eurodollar futures contracts in the 1990s. This is exactly the same
data as used by Bouchaud \cite{data1} as well as Baaquie and Srikant
\cite{Baaqmar1}. In Bouchaud \cite{data1}, the spread of the forward
rates and the eigenfuctions of its changes in time are analyzed. For
our purposes, we found it more useful to look at the scaled
multivariate cumulants of the changes in forward rates for different
maturity times.

\section{Assumptions behind the tests of the models}
The main assumption that has to be made for all the tests of the
models is that of time translation invariance. In other words, we have
to assume that $\sigma(t, \theta)$ is actually only dependent on
$\theta$ and not explicitly on $t$. We also assume that the propagator
$D(\theta, \theta')$ has no explicit time dependence which is possible
in principle. It is reasonable and conceptually economical to assume
that different times in the future are equivalent. Further, carrying
out any meaningful analysis while these quantities are subject to
changes in time is impossible.

\begin{figure}[h]
  \centering
  \epsfig{file=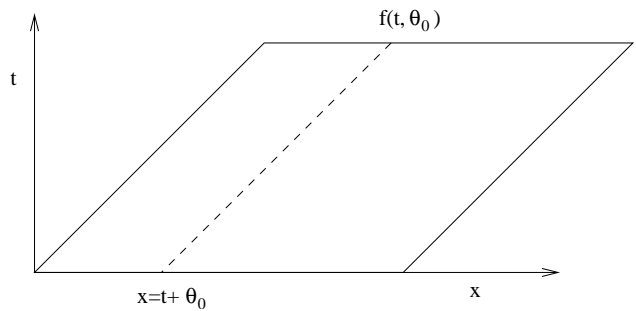, height=4cm}
  \caption{The lines of constant $\theta$ for which we have obtained
    the forward rates by linear interpolation from the actual forward
    rates which are specified at constant $x$. } 
  \label{fig:dataf}
\end{figure}

Another important assumption that has to be made is that the forward
rate curve is reasonably smooth at small intervals at any given point
in time. This assumption is very difficult to test in any meaningful
sense given the relative paucity of data as forward rate data is
available only at 3 month intervals (which is what necessitates this
assumption in the first place). However, the assumption is a
reasonable one to make as one would intuitively expect that the
forward rate, say three years into the future would not be too
different from that three years and one month into the future. In
fact, we will show later that there seems to be strong evidence of
very long term correlations in the movements of the forward rate. This
seems to make the smoothness assumption reasonable as nearby forward
rates tend to move together (except possibly at points very close to
the current time). This assumption is required as the forward rate
data is provided for constant maturity which we have been denoting by
$x$ while we want data for constant $\theta$, as shown in figure
\ref{fig:dataf}. With this assumption, we can get the data by simple
linear interpolation. The loss in accuracy due to this linear
interpolation is not all that serious if $\epsilon$, the time interval
of $t$ between specifications of the forward rates is small as the
random changes which we are interested in will be much larger than the
introduced errors. This same procedure was used in Matacz and Bouchaud
\cite{data1} as well as Baaquie and Srikant \cite{Baaqmar1}. 

\begin{figure}[h]
  \centering
  \epsfig{file=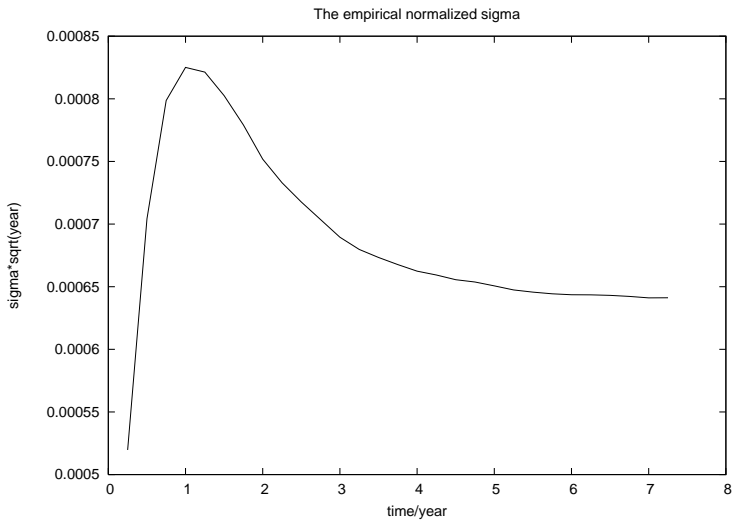, width=7cm, angle=-90}
  \caption{The empirically determined function $\sigma(\theta)$.}
  \label{fig:sigma}
\end{figure}

\section{The Correlation Structure of the Forward Rates}
A very interesting quantity to look at in the analysis of forward
rates $f(t, \theta)$ is the correlation (or scaled covariance) among
their changes for different $\theta$. Specifically we are interested
in the correlation between $\delta f(t, \theta)$ and $\delta f(t,
\theta')$, where $\delta f(t, \theta) = f(t+\epsilon, \theta) - f(t,
\theta)$. Using a free (Gaussian) quantum field theory model, this quantity should be equal to
\begin{equation}
\label{correlator}
{\cal C}(\theta, \theta') = \frac{\langle\delta f(t, \theta) \delta
  f(t, \theta')\rangle - \langle\delta f(t, \theta)\rangle \langle
  \delta f(t, \theta') \rangle}{\sqrt{\langle\delta f^2(t,
    \theta)\rangle - \langle\delta f(t, \theta)\rangle^2}
  \sqrt{\langle\delta f^2(t, \theta')\rangle - \langle \delta f(t,
    \theta')\rangle^2}} = \frac{D(\theta, 
  \theta')}{\sqrt{D(\theta, \theta) D(\theta', \theta')}}
\end{equation}
To a reasonable degree of accuracy, we can ignore the first order
expectations such as $\langle \delta f(t,\theta) \rangle$ as they are
much smaller than the second order expectations if $\epsilon$ is
small. For an $\epsilon$ of one day, the error is completely
negligible especially given the other approximations. We will do so
for the rest of the chapter. If we have a model for the propagator
$D(\theta, \theta')$, we have a prediction for this correlation
structure. Alternatively, we can use the correlation structure to fit
free parameters in $D(\theta, \theta')$. 

\begin{figure}[h]
  \begin{center}
    \epsfig{file=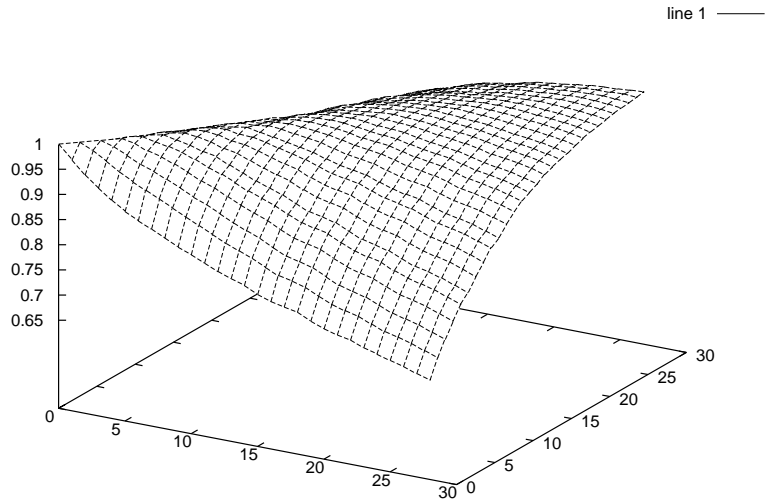, width=8cm, angle=-90}
    \caption{The correlation structure observed in the market.}
    \label{fig:market_corr}
  \end{center}
\end{figure}

It should be noted that for free (Gaussian) quantum fields the
correlation is {\em independent} of $\sigma(\theta)$, so no assumption
of its form has to be made. This is the reason why we used the scaled
covariance rather than the covariance itself to perform the study. It
is equivalent to fixing the inherent freedom in the quantities
$\sigma$ and $D$ to make $D(\theta, \theta) = 1$. The reduction in the
freedom of $\sigma$ also allows us to directly estimate it from data
since we have $\sigma(\theta) = \sqrt{<\delta f^2(t,\theta)>}$ if
$D(\theta, \theta)=1$. This is shown in figure \ref{fig:sigma}.
Further, the correlation between innovations in the forward curve is
given exactly by $D$. The correlation structure in the market
estimated from the Eurodollar futures data is shown in figure
\ref{fig:market_corr}. The structure is fairly stable in the sense
that the correlation structure for different sections of the data are
reasonably similar.

Since the propagator is always symmetric, it will be convenient to
calculate only $D(\theta_{<}, \theta_{>})$ for the different models
where $\theta_{<} = \min(\theta, \theta')$ and $\theta_{>} =
\max(\theta, \theta')$ for purposes of comparison. 

\begin{figure}[h]
  \centering
  \epsfig{file=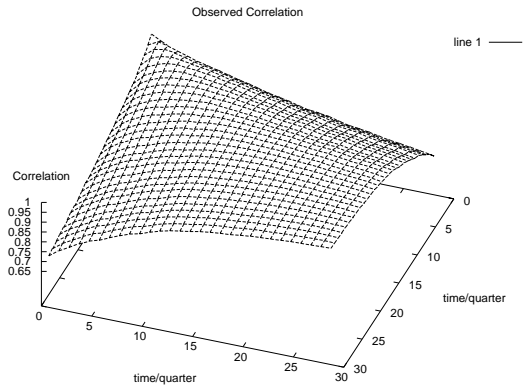, angle=-90, width=8cm}
  \caption{Observed correlation structure : a different view.}
  \label{fig:market_corr_one}
\end{figure}

\begin{figure}[h]
  \centering
  \epsfig{file=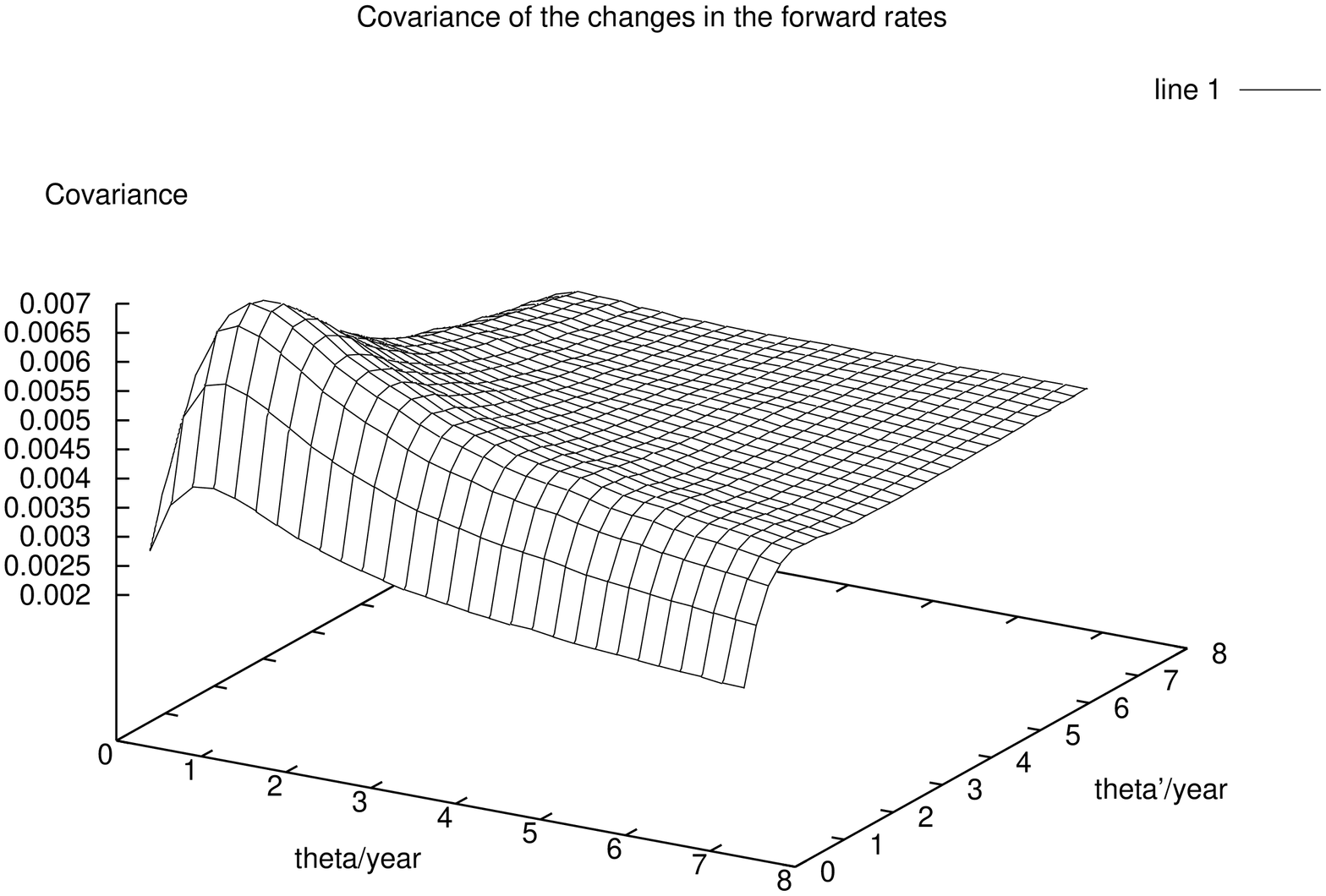, width=8cm, angle=-90}
  \caption{The covariance of innovations of forward rates observed in the market}
  \label{fig:market_covar}
\end{figure}

For the one factor HJM model, this correlation structure is constant
as all the changes in the forward rates are perfectly correlated. In
other words, $D(\theta, \theta') = 1$. For the two factor HJM model,
the predicted correlation structure is given by 
\begin{equation}
  \label{eq:twohjm}
  {\cal C}(\theta, \theta') = \frac{\sigma_1(\theta)\sigma_1(\theta') +
      \sigma_2(\theta)\sigma_2(\theta')}{\sqrt{\sigma_1^2(\theta)+\sigma_2^2(\theta)} \sqrt{\sigma_1^2(\theta') + \sigma_2^2(\theta')}} = \frac{1 + g(\theta)g(\theta')} {\sqrt{1+g^2(\theta)}\sqrt{1+g^2(\theta')}}
\end{equation}
We see that this correlation structure depends on a function of
$g(\theta) = \frac{\sigma_1(\theta)}{\sigma_2(\theta)}$. Hence, a whole function
has to be fitted from the correlation structure, something which is
quite infeasible. The covariance might be a better quantity to test
the two factor HJM model as the prediction of the covariance has a
simpler form 
\begin{equation}
  \label{eq:covarhjm}
  C(\theta, \theta') = \sigma_1(\theta) \sigma_1(\theta') +
  \sigma_2(\theta)\sigma_2(\theta') 
\end{equation}
We still need to specify a functional form for $\sigma_1$ and
$\sigma_2$ as it is not possible to estimate entire functions from
data. The usual specification of $\sigma_1(\theta) = \sigma_0$ and $
\sigma_2(\theta) = \sigma_1 e^{-\lambda \theta}$ inspired by the
assumption that the spot rate follows a Markov process is easily seen
to be unable to explain many features of the covariance in figure
\ref{fig:market_covar} such as the peak at one year or the sharp
reduction in the covariance as the maturity goes to zero. We can
straightaway conclude that the one factor HJM model is insufficient to
characterize the data while the two factor HJM model provides us with
too much freedom as we can put in an entire arbitrary function to
explain the correlation structure. If we try to reduce the freedom by
theoretical considerations, we are again unable to explain the data.

We will see that the field theory model with constant rigidity, while
explaining some features of the correlation, does not predict the
correlation very well. Hence, we consider generalizations to the
constant rigidity model. 

\section{Analysis of Field Theory Model with Constant Rigidity}
We have analysed this model in detail in the previous section. We have
seen that the model describes the innovations in the forward rates in
terms of a Gaussian random field $A$ whose structure is defined by the
action in (\ref{eq:adef}). For convenience, we repeat the action below
in terms of the variables $t$ and $\theta=x-t$ 
\begin{equation}
  \label{eq:uncons_rig_action}
  S = -\frac{1}{2} \int_{t_0}^{t_1} dt \int_0^\infty d\theta \left(A^2 +
  \left(\frac{\partial A}{\partial \theta}\right)^2\right)
\end{equation}

\begin{figure}[h]
  \centering
  \epsfig{file=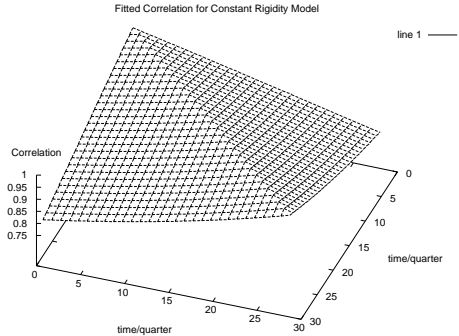, angle=-90, width=7cm}
  \caption{Fitted correlation for constant rigidity model} 
  \label{fig:cons_rig_corr}
\end{figure}

To obtain the predicted correlation structure from the propagator
(\ref{eq:D}), we have to take the limit $T_{FR} \rightarrow
\infty$ and obtain
\begin{equation}
  D(\theta, \theta') = \mu e^{-\mu \theta_{>}} \cosh \mu \theta_{<} =
  \frac{\mu}{2} \left(e^{-\mu |\theta - \theta'|} + e^{-\mu (\theta +
      \theta')}\right)
\end{equation}
The predicted correlation structure for this model can be found from
this form of the propagator by normalization and from
(\ref{correlator}) is given by
\begin{equation}
  \label{eq:corr_uncons}
  {\cal C}(\theta, \theta') = \sqrt{\frac{e^{-\mu
        \theta_{>}} \cosh \mu \theta_{<}}{e^{-\mu \theta_{<}} \cosh
      \mu \theta_{>}}}
\end{equation}
when the limit $T_{FR} \rightarrow \infty$ is taken.  To estimate the
parameter $\mu$ from market data, we use the Levenberg-Marquardt
method from Press et al \cite{NumRecipes} to fit the parameters to the
observed correlation structure graphed in figure
\ref{fig:market_corr}. The fitting was done by minimizing the square
of the error. The overall correlation was fitted by $\mu =
0.061/$year. To obtain the error bounds, the data was split into 346
data sets of 500 contiguous days of data each and the estimation done
for each of the sets. The 90\% confidence interval for this data set
is $(0.057, 0.075)$. Note that the confidence interval is asymmetric
from the overall best fit due to the nonlinear dependence of the
correlation (\ref{eq:corr_uncons}) on $\mu$. The root mean square for
the correlation for the best fit value is 4.23\% which shows that the
model's prediction for the correlation structure is not very good. The
main problem as can be seen from a comparison between the prediction
for the best fit $\mu$ in figure \ref{fig:cons_rig_corr} and the
actuall correlation structure in figure \ref{fig:market_corr} is that
the prediction is largely independent of the actual value of $\theta$
and largely determined by $|\theta - \theta'|$ which is not the case
in reality. The correlation rapidly increases as $\theta$ increases in
reality.

\begin{figure}[h]
  \centering
  \epsfig{file=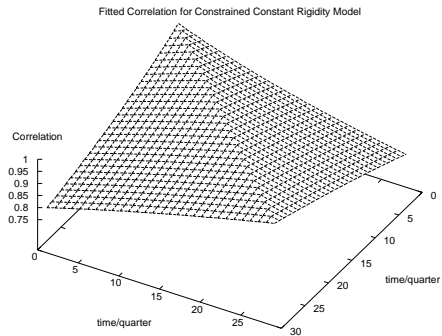, angle=-90, width=7cm}
  \caption{Fitted correlation for constrained field theory model}
  \label{fig:constrained_prop}
\end{figure}

\section{Field Theory with Constrained Spot Rate}

One clear fact we notice from the covariance of the innovations in the
forward rates in figure \ref{fig:market_covar} is that the covariance
falls rapidly as $\theta \rightarrow 0$. This observation leads one to
a model \cite{Baaquienotes} where $A(t,0)$ is constrained to follow a
normal distribution with variance $a$. The mean of $A(t,0)$ can be
fixed at any value but will cause a corresponding change in
$\alpha(0)$ which makes the mean value irrelevant.  For calculational
purposes it is easiest to assume that it remains at zero. This
constraint can be implemented by modification of the action to
\begin{equation}
  \label{eq:cons_rig_action}
  e^{S_{constrained}} = \int_{-\infty}^{\infty} d\xi S e^{i\xi
    A(t,0)}e^{-a^2\xi^2/2} 
\end{equation}
where $S$ is the action specified in (\ref{eq:uncons_rig_action}). The
propagator $D(\theta, \theta')$ for this model is given by
\begin{equation}
  \label{eq:prop_cons}
  D(\theta, \theta') = \mu e^{-\mu \theta_{>}}\left(\cosh \mu \theta_{<} -
    \frac{\mu e^{-\mu \theta_{<}}}{\mu+a}\right)
\end{equation}
After normalizing, we see that the prediction for the correlation
structure is given by 
\begin{equation}
  \label{eq:corr_cons}
  {\cal C}(\theta, \theta') = 
  \sqrt{\frac{e^{-\mu \theta_{>}} (\cosh \mu \theta_{<} - \frac{\mu e^{-\mu
          \theta_{<}}}{\mu + a})}{e^{-\mu \theta_{<}}(\cosh \mu
      \theta_{>} - \frac{\mu e^{-\mu \theta_{>}}}{\mu + a})}}
\end{equation}

We can see that the free parameters are $\mu$ and $a$. Further, it
will be seen that it is easier to consider the ratio $a/\mu^2$ as it
is dimensionless. The results of the Levenberg-Marquardt method showed
that the fitted value of $\mu$ and $a$ were very small, of the order
of $10^{-7}$/year for $\mu$ and $10^{-13}/\mathrm{year}^2$ for $a$
both being very unstable but the ratio $a/\mu^2$ was stable with a
value in the range $(6.7,10.7)$ with an overall best fit of $9.4$. The
most reasonable explanation for this behaviour is that the ratio
$a/\mu^2$ determines the behaviour of (\ref{eq:corr_cons}) for small
$\mu$ and it is this region of parameter space that gives a
correlation structure closest to the empirically observed one. We see
from the fitted propagator in figure \ref{fig:constrained_prop} that
the behaviour at large $\theta$ is slightly better when the constraint
is put in. The root mean square error was 3.35\% which again means the
fit was not very good though significantly better than if the
constraint was not applied.  It must be recognized that the constraint
introduces one extra free parameter which should improve the best fit.
Hence, we see that this model, while again performing better than HJM,
is still not very accurate.  While the results are not very good, they
do represent a reasonable first approximation and are still
significantly better than the one factor HJM model.

\section{Field Theory Model with Maturity Dependent Rigidity
  $\mu=\mu(\theta)$} 
Another way to get a correlation structure that depends directly on
the values of $\theta$ and $\theta'$ in a significant way and not only
on their difference is to make $\mu$ a function of $\theta$. This has
a direct physical meaning as it means that if we imagine the forward
rate curve as a string, its rigidity is increasing as maturity
increases making the $A$ for larger $\theta$ more strongly correlated
if $\mu$ decreases as a function of $\theta$. We choose the function
$\mu = \frac{\mu_0}{1+\lambda \theta}$ as it declines to zero as
$\theta$ becomes large as is expected from the observed covariance in
figure \ref{fig:market_covar}, contains the constant $\mu$ case as a
limit and is solvable. The action is given by
\begin{equation}
  \label{eq:noncons_rig_action}
  S = -\frac{1}{2} \int_{t_0}^{t_1} dt \int_0^\infty d\theta \left(A^2
    + \left(\frac{1+\lambda \theta}{\mu_0} \frac{\partial
        A}{\partial \theta}\right)^2\right)
\end{equation}
This is still a quadratic action and can be put into a quadratic form
by performing integration by parts and setting the boundary term to
zero since we are assuming Neumann boundary conditions. The inverse
(Greens function) of the quadratic operator or the propagator for this
action is found to be 
\begin{equation}
  \label{eq:prop_linear}
  \begin{split}
    D(\theta, \theta'; T_{FR}) =& \frac{\mu_0^2 \alpha}{2\lambda
      \alpha (\alpha+1/2) (1-(1+\lambda 
      T_{FR})^{-2\alpha})}\\
    &\left(\frac{\alpha+1/2}{\alpha-1/2} (1 +
      \lambda T_{FR})^{-2\alpha} (1+\lambda \theta_{>})^{\alpha-1/2} +
      (1+\lambda \theta_{>})^{-\alpha-1/2}\right)\\
    &\left(\frac{\alpha+1/2}{\alpha-1/2} (1 +
      \lambda T_{FR})^{-2\alpha} (1+\lambda \theta_{<})^{\alpha-1/2} +
      (1+\lambda \theta_{<})^{-\alpha-1/2}\right)
  \end{split}
\end{equation}
where $\alpha = \sqrt{\frac{1}{4} + \frac{\mu_0^2}{4 \lambda^2}}$ and
where we have put the bound on the $\theta$ variable $T_{FR}$
explicitly. The reason for this is that the limits have to be taken
carefully in order to compare this model to the HJM in the limit
$\mu_0 \rightarrow 0$ and to the constant rigidity field theory model
when $\lambda \rightarrow 0$.

Let us first consider the limit $\lambda \rightarrow 0$. First, we
note 
\begin{equation}
  \label{eq:limit1}
  \alpha = \sqrt{\frac{1}{4} + \frac{\mu_0^2}{\lambda^2}} \sim
  \frac{\mu_0}{\lambda} \sqrt{1 + \frac{\lambda^2}{4 \mu_0^2}}
  \sim \frac{\mu_0}{\lambda} 
\end{equation}
Therefore, we have 
\begin{equation}
  \label{eq:limit2}
  (1 + \lambda \theta)^{-\alpha - 1/2} = \left((1 + \lambda
    \theta)^{1/\lambda}\right)^{-\mu_0} (1 + \lambda \theta)^{-1/2}
  \sim e^{-\mu_0 \theta}
\end{equation}
Similarly $(1+\lambda \theta)^{\alpha -1/2} \sim e^{\mu_0\theta}$,
$(1+\lambda \theta)^{-\alpha - 1/2} \sim e^{-\mu_0 \theta}$ and
$(1+\lambda T_{FR})^{-2\alpha} \sim e^{-2\mu_0 T_{FR}}$. Putting all
these limits into (\ref{eq:prop_linear}) and performing some
straightforward simplifications, we see that (\ref{eq:prop_linear})
becomes equal (\ref{eq:D}) in the limit $\lambda \rightarrow 0$. In the
taking of this limit, we did not have any trouble with
$T_{FR}$. However, for the HJM limit, we will see that the limit
$T_{FR} \rightarrow \infty$ has to be taken only after the limit
$\mu_0 \rightarrow 0$ has been taken. 

Let us now consider the limit $\mu_0 \rightarrow 0$.  In this limit
$\alpha \sim \frac{1}{2} + \frac{\mu_0^2}{\lambda^2}$. Hence, only one term
in (\ref{eq:prop_linear}) survive as all the others are multiplied by
$\alpha - 1/2$. This surviving term can be evaluated
\begin{equation}
  \label{eq:limit3}
  \begin{split}
    &\frac{\mu_0^2}{2\lambda} \frac{(\alpha+1/2)^2}{\alpha - 1/2}
    \frac{1}{1 - (1 + \lambda T_{FR})^{-1}} (1 + \lambda T_{FR})^{-1}\\
    &= \frac{\mu_0^2}{2\lambda} \times \frac{2\lambda^2}{\mu_0^2} \times
    \frac{1+\lambda T_{FR}}{\lambda T_{FR}} \times
    \frac{1}{1+\lambda T_{FR}}\\
    &= \frac{1}{T_{FR}}
    \end{split}
\end{equation}
The terms $(1+\lambda \theta_{>})^{\alpha-1/2}$ and $(1 + \lambda
\theta_{<})^{\alpha - 1/2}$ obviously go to one in this limit and so
were not included in the calculation above. This result can be seen to
be equivalent to the HJM propagator after normalization. If the limit
$T_{FR} \rightarrow \infty$ is taken first, then the propagator
becomes
\begin{equation}
  \label{eq:limit4}
  D(\theta, \theta') = \frac{\mu_0^2(\alpha - 1/2)}{2\lambda\alpha
    (\alpha + 1/2)} (1 + \lambda \theta_{>})^{-\alpha -1/2}
  \left(\frac{\alpha + 1/2}{\alpha - 1/2} (1 + \lambda
    \theta_{<})^{\alpha - 1/2} + (1 + \lambda \theta_{<})^{-\alpha -
      1/2}\right)
\end{equation}
which exhibits a $\theta$ dependence in the limit $\mu_0 \rightarrow
0$. Hence, this cannot be made equivalent to HJM if the limits are
taken in the wrong order. This problem is not present in the constant
rigidity model.

For comparison with market data, we still take the limit $T_{FR}
\rightarrow \infty$ as the model is then still directly related to the
field theory model. The predicted correlation structure for this model
is then given by
\begin{equation}
  \label{eq:corr_ncrig}
  {\cal C}(\theta, \theta') = \left(\frac{(\alpha + 1/2) (1+\lambda
      \theta_{<})^{2\alpha} + \alpha 
        - 1/2} {(\alpha + 1/2) (1+\lambda \theta_{>})^{2\alpha} +
        \alpha - 1/2}\right)^{\frac{1}{2}}
\end{equation}

\begin{figure}[h]
  \centering
  \epsfig{file=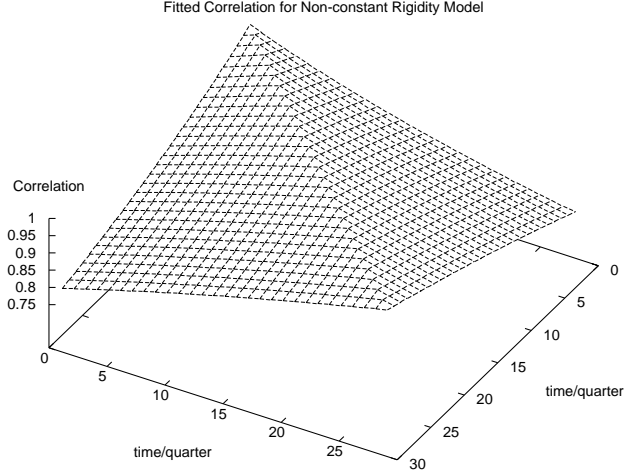, width=7cm, angle=-90}
  \caption{Fitted correlation structure for the non-constant rigidity model.}
  \label{fig:corr_ncrig}
\end{figure}

We fitted the parameters $\mu_0$ and $\lambda$ to the correlation
structure observed in the market in a similar manner as for the field
theory model and obtained the results $\mu_0 = 1.2 \times
10^{-5}/\mathrm{year}$ and $\lambda = 0.108/\mathrm{year}$. The root
mean square error in the correlation was 3.35\%. On performing the
error analysis for the parameters, it is found that $\mu_0$ is very
unstable but always very small (less than $10^{-2}/\mathrm{year}$)
while the 90\% confidence interval for $\lambda$ is (0.099, 0.149).
The relatively high value for $\lambda$ seems to show that the falloff
of the rigidity paramater $\mu = \frac{\mu_0}{1+\lambda \theta}$ is
fairly rapid. The error is reduced from 4.23\% to 3.35\% but an extra
parameter has had to be added and the model has become considerably
more complicated due to the freedom of the form of the rigidity
parameter $\mu$. Further, we seem to be in the region of very small
$\mu_0$ which does not behave well in the HJM limit. In fact, the
correlation structure in this limit is given by
\begin{equation}
  \label{eq:limit5}
  {\cal C}(\theta, \theta') = \sqrt{\frac{1+\lambda \theta_{<}}{1 +
      \lambda \theta_{>}}}
\end{equation}
Due to the very small value of $\mu_0$ for the fitted function, this
is a very good approximation for the fit. The obtained fit for the
correlation function can be seen in figure \ref{fig:corr_ncrig}. 

The limited improvement, the relatively complicated form of the
correlation and the near zero $\mu_0$ problem prompted us to consider
a different way of approaching the problem which presented a much more
satisfactory solution. This model is described in the next section

\section{Field Theory Model with $f(t,z(\theta))$}
To see where we might make an improvement, we notice that the
predicted correlation structure with the field theory model is largely
defined by the $e^{-\mu|\theta_{>} - \theta_{<}|}$ term which means
that the correlation does not depend explicitly on the times
$\theta_{>}$ and $\theta_{<}$\footnote{There is another term of the
  form $e^{-\mu(\theta_{>} + \theta_{<})}$ but this has only a small
  effect on the correlation structure}.  However, we see immediately
from figure \ref{fig:market_corr} that the correlation increases
significantly as we increase $\theta_{>}$ and $\theta_{>}$. This is
intuitively reasonable as market participants are likely to treat the
difference between ten and fifteen years into the future quite
differently from the difference between now and five years. Far out
into the future, we would expect all times to be equivalent. In other
words, there is good reason to expect $\lim_{\theta_{<} \rightarrow
  \infty} D(\theta_{>}, \theta_{<}) = 1$ \footnote{Obviously,
  $\theta_{<} \rightarrow \infty$ automatically implies $\theta_{>}
  \rightarrow \infty$}. This is not satisfied by the constant rigidity
models or by the varying rigidity model (if the limit $T_{FR}
\rightarrow \infty$ is taken). For the latter model, this is slightly
surprising since $\mu \rightarrow 0$ as $\theta \rightarrow \infty$
and we might expect that for large $\theta$ the varying rigidity model
should go into the HJM model limit ($D=1$). However, this does not
happen as previously discussed since we have taken the limit $T_{FR}
\rightarrow \infty$.

Further, the relatively marginal reduction of the error shows that
varying the rigidity parameter does not quite reflect the data. An
alternative way to consider the problem would be to use the observed
correlation structure to induce a metric onto the $\theta$ direction.
In some sense, this metric would be measuring the ``psychological
distance'' in the investor's minds which corresponds to a certain
separation in maturity time. To make this concrete, let us write the
observed correlation as $D(\theta, \theta') = e^{-s(\theta,
  \theta')}$. Since $D(\theta, \theta)=1$, $s(\theta, \theta)=0$ and
$s$ is symmetric as well. If we can show the triange property (which
in the case of one dimension reduces to the straightforward condition
that $s(\theta_1, \theta_3) = s(\theta_1, \theta_2) + s(\theta_2,
\theta_3)$), we can see that $s$ makes a good definition of distance in
$\theta$. From the market data, it can be shown that this rule is
very approximately satisfied and we can use it as an approximate way to
induce a metric onto the $\theta$ direction from the observed market
data.

It should also be noted that introducing the metric is different from
changing the form of the rigidity function $\mu(\theta)$. To see this,
we write the action with the rigidity function $\mu(\theta)$ as 
\begin{equation}
  S_{old} = -\frac{1}{2}\int_{t_0}^{t_1} dt \int_0^\infty d\theta \left(A^2
  + \frac{1}{\mu^2}\left(\frac{\partial A}{\partial z}\right)^2\right)
\end{equation}
where the functional variation of $\mu$ with $\theta$ has been
absorbed into the variable $z = g(\theta)$ (where $g$ is invertible)
so that the $\mu$ above is a constant.  With a change of variables we
get the action as
\begin{equation}
  S_{old} = -\frac{1}{2}\int_{t_0}^{t_1} dt \int_{g(0)}^{g(\infty)} dz
  h'(z) \left(A^2 + \frac{1}{\mu^2}\left(\frac{\partial A}{\partial z}\right)^2\right) 
\end{equation}
where $h=g^{-1}$. With the introduction of the metric, we obtain the
action
\begin{equation}
  S_{new} = -\frac{1}{2} \int_{t_0}^{t_1} dt \int_{g(0)}^{g(\infty)} dz 
  \left(A^2 + \frac{1}{\mu^2} \left(\frac{\partial A}{\partial
        z}\right)^2\right) \ne S_{old} 
\end{equation}

The Green's functions for $S_{new}$ should be solved using the $z$
variables, and as expected the solution is given by $D(z,z') =
\frac{1}{2}(\exp{-\mu|z-z'|} + \exp{-\mu(z+z')})$. It can be shown
that the martingale condition is satisfied with the Green's function
given by $D(z,z')$.

\begin{figure}[h]
  \centering
  \epsfig{file=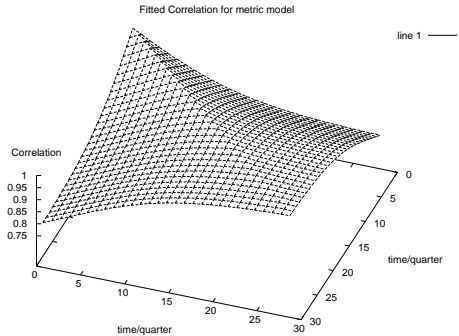, angle=-90, width=7cm}
  \caption{Fitted correlation for the model with metric $g(\theta) = \tanh \beta \theta$}
  \label{fig:metric_corr}
\end{figure}

Bearing in mind the condition that, at large $\theta$, the
correlations should be close to 1, or equivalently that the distance
should be small, we choose a metric that satisfies property,
$g(\theta) = \tanh \beta \theta$. We use this form of the metric to
fit the correlation structure and obtain the result that $\mu =
0.48/\mathrm{year}$ and $\beta = 0.32/\mathrm{year}$ with a root mean
square error of only 2.46\%. Both parameters are also stable when the
error analysis for the parameters is carried out. The 90\% confidence
interval for $\mu$ is (0.45, 0.58) and that for $\beta$ is (0.22,
0.33). Hence, we see that even the parameter estimation for this model
is more robust as the parameters are atleast stable. Further, the
shape of the fitted function is clearly closer to the observed one as
can be seen from figures \ref{fig:market_corr_one},
\ref{fig:cons_rig_corr}, \ref{fig:constrained_prop},
\ref{fig:corr_ncrig} and \ref{fig:metric_corr}. The error that remains
is largely confined to the correlation between the spot rate and other
forward rates which is not too surprising since the spot rate behaves
very differently from the other forward rates.

We emphasize here that this involves a fundamentally new way of
thinking of the interest rate models. So far, we have made models
which generalized HJM so as to achieve a theory without too little
freedom as in the one factor HJM model or too much freedom as in the
two factor HJM model. While retaining this framework, we now use
empirical data to guide us in refining the model to give us an insight
into market psychology which will result from the induced metric. 

\section{Acknowledgements}
We would like to thank Jean-Philippe Bouchaud and Science and Finance
for kindly providing us with the data used for this study.

\bibliography{list-int}

\begin{thebibliography}{1}

\bibitem{Baaquie}
B.~E. Baaquie,
\newblock Physical Review E {\bf 64}, 1 (2001).

\bibitem{Baaquiesv}
B.~E. Baaquie,
\newblock Physical Review E {\bf 65}, 056122 (2002),
\newblock cond-mat/0110506.

\bibitem{Baaquienotes}
B.~E. Baaquie,
\newblock Quantum finance,
\newblock In preparation, 2002.

\bibitem{data1}
A.~Matacz and J.-P. Bouchaud,
\newblock International Journal of Theoretical and Applied Finance {\bf 3}, 703
  (2000).

\bibitem{Baaqmar1}
B.~E. Baaquie and M.~Srikant,
\newblock Empirical {I}nvestigation of a {Q}uantum {F}ield of {F}orward
  {R}ates,
\newblock National University of Singapore
  http://xxx.lanl.gov/abs/cond-mat/0106317, 2002.

\bibitem{NumRecipes}
W.~H. Press, S.~A. Teukolsky, W.~T. Vettering, and B.~P. Flannery,
\newblock {\em Numerical Recipes in C : The Art of Scientific Computing},
\newblock Cambridge University Press, 1995.

\end{thebibliography}
\bibliographystyle{phaip}
\end{document}